\documentclass[runningheads]{llncs}

\usepackage[T1]{fontenc}
\usepackage[utf8]{inputenc}
\usepackage{lmodern}
\usepackage{microtype}
\usepackage{amsmath,amssymb}
\usepackage{graphicx}
\usepackage{booktabs}
\usepackage{array}
\usepackage{multirow}
\usepackage{enumitem}
\usepackage[hidelinks]{hyperref}
\usepackage{placeins}

\setlist{nosep,leftmargin=*}

\title{Optimization Geometry of QAOA and Variational Quantum Algorithms}
\titlerunning{Optimization Geometry of QAOA and Variational Quantum Algorithms}

\author{%
Vojt\v{e}ch Nov\'{a}k\inst{1,2,3}\thanks{Corresponding author.} \and
Ivan Zelinka\inst{1,2,3} \and
Silvie Ill\'{e}sov\'{a}\inst{4} \and
Swagatam Das\inst{5} \and
Martin Beseda\inst{6}}
\authorrunning{V. Nov\'{a}k et al.}

\institute{%
Department of Computer Science, Faculty of Electrical Engineering and Computer Science,
VSB -- Technical University of Ostrava, Ostrava, Czech Republic\\
\email{vojtech.novak.st1@vsb.cz}
\and
IT4Innovations National Supercomputing Center, VSB -- Technical University of Ostrava,
708 00 Ostrava, Czech Republic
\and
Department of Informatics and Statistics, Marine Research Institute, Klaipeda University,
Klaipeda, Lithuania
\and
Gran Sasso Science Institute, L'Aquila, Italy
\and
Electronics and Communication Sciences Unit, Indian Statistical Institute,
700108 Kolkata, India
\and
Dipartimento di Ingegneria e Scienze dell'Informazione e Matematica,
Universit\`{a} dell'Aquila, Via Vetoio, I-67010 Coppito, L'Aquila, Italy}

\begin{document}
\maketitle

\begin{abstract}
Variational quantum algorithms turn choices of Hamiltonian, ansatz, and parameterization into a classical nonconvex optimization problem. We study how this objective function can be visualized and characterized in ways that help explain optimizer behavior. We distinguish two properties of the objective: the number of local minima encountered along sampled directions and the differences in quality among local-search endpoints. We then ask how a local optimizer, represented by BFGS, compares with adaptive differential evolution, represented by jSO. Rather than comparing jSO with a single local run, we allow BFGS multiple starts within the same function-evaluation budget. This gives local search repeated opportunities to explore different basins and provides a stronger baseline for asking when a global evolutionary solver is useful.

We study these questions using VQE and QAOA, focusing on how frustration, circuit depth, parameter tying, mixed locality, and nonlinear reparameterization change the Hamiltonian expectation-value objective seen by the classical optimizer. Our analysis combines exact low-dimensional restrictions, optimizer-informed views, basin summaries, classical optimization analogies, and simple empirical diagnostics.

Increasing independent circuit depth raises the sampled local-minimum count without making global search more effective. Parameter tying, by contrast, produces both more repeated local structure and much larger differences in quality among local-search endpoints, and in this regime adaptive differential evolution outperforms function-evaluation-matched multistart BFGS. The comparison shows that the number of local minima alone does not determine whether global search is advantageous: the important distinction is whether different basins lead to similarly good solutions or to substantially different objective values. These results provide a practical workflow for connecting model construction, objective-function geometry, empirical diagnostics, and optimizer choice.
\end{abstract}

\section{Introduction}
\label{sec:intro}

Variational quantum algorithms (VQAs) such as VQE and QAOA are hybrid procedures: a quantum
circuit produces objective values, while a classical optimizer decides which circuit parameters to try
next~\cite{peruzzo2014,farhi2014,zhou2020,cerezo2021,tilly2022}. Because of this hybrid structure, the
practical difficulty of a VQA depends not only on the quantum circuit or the Hamiltonian separately,
but also on the classical objective function presented to the optimizer,
\begin{equation}
E(\boldsymbol{\theta}) =
\langle 0|U^{\dagger}(\boldsymbol{\theta}) H U(\boldsymbol{\theta})|0\rangle
=
\langle \psi(\boldsymbol{\theta}) | H | \psi(\boldsymbol{\theta}) \rangle,
\label{eq:objective}
\end{equation}
where $|\psi(\boldsymbol{\theta})\rangle = U(\boldsymbol{\theta})|0\rangle$ is the prepared quantum state.
In quantum physics, this is the familiar \emph{Hamiltonian expectation value}. In linear algebra and
optimization, the mapping $|\psi\rangle \mapsto \langle \psi|H|\psi\rangle$ is a Hermitian
\emph{quadratic form}, often written as the \emph{Rayleigh quotient} for normalized state vectors:
\begin{equation}
R_H(|\psi\rangle) =
\frac{\langle \psi|H|\psi\rangle}{\langle \psi|\psi\rangle}.
\label{eq:rayleigh-quotient}
\end{equation}
The circuit parameterization defines the map
$\boldsymbol{\theta} \mapsto |\psi(\boldsymbol{\theta})\rangle$.
Evaluating the physical energy on the states reachable through this map produces the classical
function $E(\boldsymbol{\theta})$ that the optimizer must navigate.

Viewed in this way, VQA optimization belongs to a broader class of nonconvex optimization problems
in which model construction determines both the set of representable solutions and the geometry
exposed to the optimizer. This creates a natural connection to familiar questions in classical
optimization: how many local minima does the objective contain, how different are those minima in
quality, and when is repeated local search sufficient compared with a more global search strategy?

These questions matter because two circuits targeting the same physical problem can present very
different optimization problems. Increasing circuit depth can increase reachability while also changing
trainability~\cite{pellowJarman2024}. Restricted parameterizations can generate repeated local traps even
when gradients remain finite~\cite{mcclean2018,anschuetz2022}, and QAOA objectives can exhibit
nontrivial basin structure~\cite{boy2024}. Parameter tying can reduce the number of free variables
while making each variable influence several circuit locations, so a lower-dimensional search space
does not necessarily imply an easier optimization problem. Similarly, a nonlinear reparameterization
can fold a simple physical motion into a more multimodal function of the coordinates seen by the
optimizer.

Related issues also appear in quantum machine learning (QML)
\cite{illesova2025classical,illesova2025complementarity,illesova2026importance,gupta2022quantum},
where parameterized quantum circuits prepare trial states and measured observables define a learning
objective or cost function. A classical optimizer then updates the circuit parameters iteratively.
Across these settings, the practical optimization problem is therefore determined jointly by the
physical model, the circuit construction, and the way the circuit is parameterized.

The aim of this paper is to give a simple visual and empirical account of how these choices shape the
objective function and, in turn, optimizer behavior. Our main operational question is:
\emph{when is local search enough, and when does broader global exploration become useful?}
Optimizer rankings are strongly problem dependent in VQAs, so a geometry-based view provides a
complement to benchmark-only comparisons
\cite{failde2023,bonet2023,novakGlobalSearch,novak2026quantum}.
Complementary studies have examined optimizer robustness under noisy VQA objectives,
Variational Hamiltonian Ansatz (VHA) specific optimization, statistical optimizer benchmarking,
and parameter efficiency
\cite{novakNoisy2026,illesovaVHA2025,illesovaStat2025,bezdekOptimization2025,novakReliable2025}.
Here we instead use small controlled examples to identify which properties of the objective are
associated with different local and global optimization behavior.

A central point of the paper is that simply calling an objective ``multimodal'' is not sufficiently
informative. We distinguish two properties. The first is the \emph{number of local minima} encountered
along sampled directions through parameter space. The second is the \emph{quality of those minima},
measured through the distribution of objective values reached by local optimization from different
starting points. These properties need not change together. An objective may contain many local minima
that all have similar quality, or it may contain minima whose objective values differ substantially.
The distinction matters because falling into the ``wrong'' minimum is much more costly in the second
case.

This distinction also motivates our optimizer comparison. We use BFGS as a representative local
optimizer and adaptive differential evolution, represented by jSO, as a representative global
evolutionary method. Importantly, BFGS is not restricted to a single run. Within the optimizer
comparison, it is restarted multiple times under the same total function-evaluation budget as jSO.
Multistart BFGS therefore has repeated opportunities to enter different attraction basins. This gives
a stronger baseline than a single local run and allows us to ask a more useful question: whether the
broader exploration of an evolutionary method still provides an advantage when local search is itself
allowed to restart.

Throughout the paper, a \emph{basin} is used operationally: it is a region of initial conditions that a
specified local-search procedure carries to the same, or an energetically equivalent, local endpoint.
Basin accessibility therefore depends on the chosen local optimizer. We do not attempt a complete
enumeration of all attraction basins of the continuous objective. Instead, we use random local starts,
sampled one-dimensional directions, and selected low-dimensional restrictions to obtain simple
optimizer-relevant summaries of the objective.

The paper studies these ideas using both QAOA and VQE. We examine how frustration, circuit depth,
parameter tying, mixed locality, and nonlinear reparameterization reshape the Hamiltonian
expectation-value objective. Our visualizations include literal coordinate slices, optimizer-informed
restrictions, basin-focused summaries, and classical optimization analogies. These views are then
combined with simple empirical diagnostics based on sampled local-minimum counts and local-search
endpoint quality.

The controlled optimizer comparison makes the motivation for separating these two properties
particularly clear. Increasing independent QAOA depth can increase the number of sampled local
minima without making local-search endpoints substantially worse. In this case, additional
multimodality alone does not make global evolutionary search more effective than function-evaluation-
matched multistart local search. Parameter tying produces a different situation: repeated local
structure is accompanied by much larger differences in the quality of local-search endpoints. Here,
ending in the wrong basin becomes materially costly, and adaptive differential evolution gains an
advantage. Thus, the number of local minima alone is not enough to decide whether a global optimizer
is useful; the quality differences between the minima reached by local search are also important.

The paper makes three contributions. First, it organizes useful visualizations of VQA objective
functions into a practical workflow that distinguishes literal slices, optimizer-informed restrictions,
and basin-focused summaries. Second, it separates directional local-minimum count from the quality
of local-search endpoints and shows how the physical model, ansatz, and parameterization can change
these properties independently. Third, it connects these visual and empirical diagnostics to a
controlled MaxCut comparison between function-evaluation-matched multistart BFGS and adaptive
differential evolution.

Section~\ref{sec:construction} describes how the physical model, ansatz, and parameterization shape the
objective function. Section~\ref{sec:workflow} presents the visualization and diagnostic tools.
Section~\ref{sec:examples} relates these measurements to optimizer behavior in QAOA, VQE, and MaxCut
examples.

\section{How model construction shapes the objective function}
\label{sec:construction}

\begin{figure}[!htbp]
    \centering
    \includegraphics[width=0.96\linewidth]{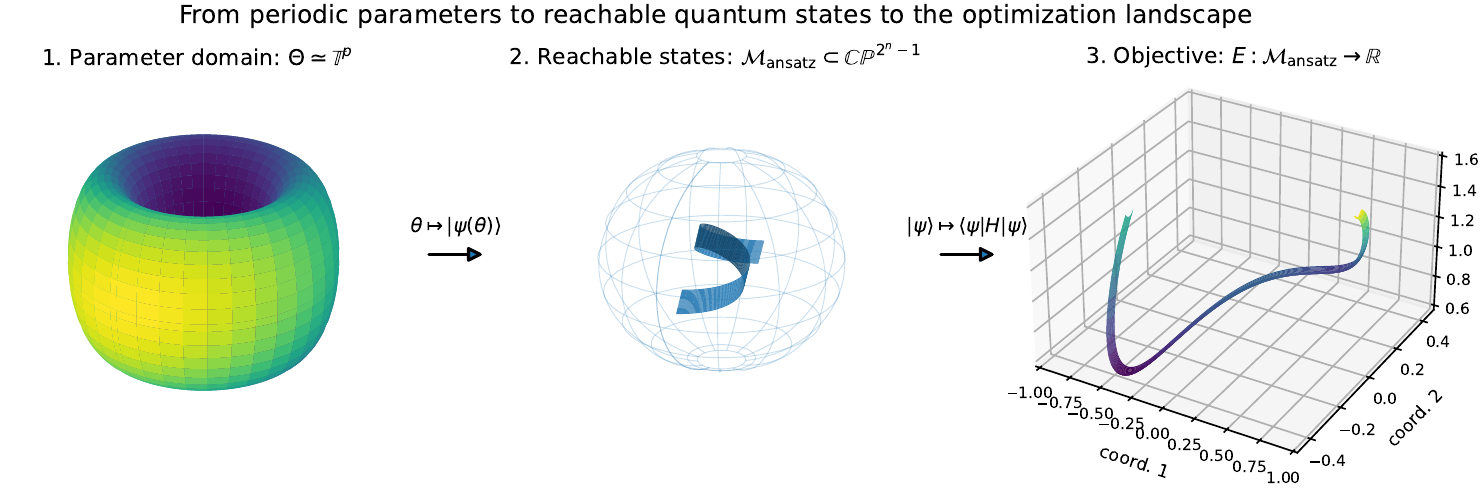}
   \caption{Conceptual overview of a variational quantum algorithm. Parameters define a circuit that prepares a restricted set of reachable quantum states, and the Hamiltonian assigns an energy to each state. The optimizer navigates the resulting classical energy landscape over the parameters.}
    \label{fig:vqa-spaces}
\end{figure}

Figure~\ref{fig:vqa-spaces} provides the paper's conceptual anchor. It separates three spaces that are
often conflated in informal discussions: the classical parameter domain, the reachable subset of
quantum states, and the scalar energy values assigned by the Hamiltonian. On the full space of normalized
states, the Rayleigh quotient Eq.~\eqref{eq:rayleigh-quotient} (or quadratic form $\langle \psi|H|\psi\rangle$) has a very simple geometry: ground-state eigenspaces are minima, intermediate excited states are saddle
points, and the top eigenspace is a maximum. In practice, however, the optimizer only explores the restricted
subset of states reachable by the chosen circuit parameterization, where additional local minima can appear.

We consider three design choices.
\begin{description}
    \item[Physical model.] Frustrated spin systems, mixed-locality couplings, and chemistry
    Hamiltonians encode different energetic preferences.
    \item[Ansatz class.] Hardware-efficient circuits, Hamiltonian-variational constructions, and
    QAOA layered circuits reach different families of states. Expressivity and entangling
    capability describe the ansatz, while optimizability also depends on the parameterization and search
    procedure~\cite{sim2019,illesovaQMetric2025}.
    \item[Parameterization.] Independent angles, parameter tying, and nonlinear coordinate maps expose
    those states to the optimizer in different ways.
\end{description}
Each lever can reshape the classical objective even when the underlying physics is fixed.

\begin{figure}[!htbp]
    \centering
    \includegraphics[width=0.98\linewidth]{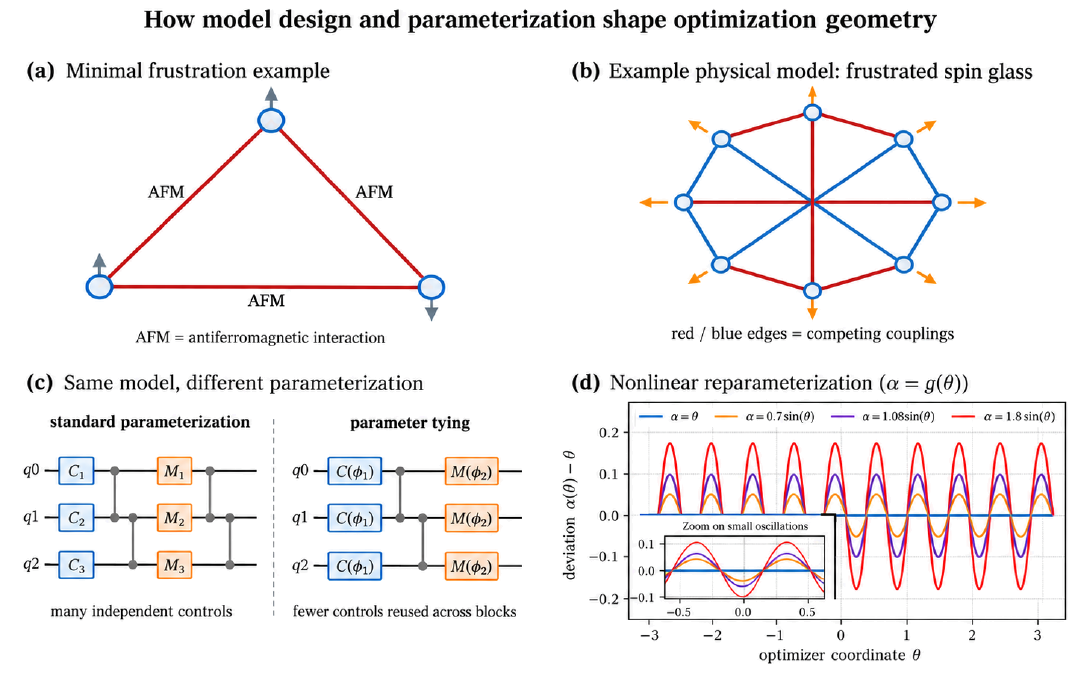}
    \caption{Compact construction overview. (a) Frustration on a triangle: not all
    antiferromagnetic preferences can be satisfied simultaneously. (b) A small frustrated spin-glass
    model used as a readable QAOA physical example. (c) The same physical model can be exposed
    through a standard parameterization with more independent controls or through parameter tying with fewer
    controls. (d) A nonlinear coordinate map can fold the variable seen by the optimizer.}
    \label{fig:construction-overview}
\end{figure}

Figure~\ref{fig:construction-overview} collects these ideas in one place. Panel~(a) introduces
frustration in the simplest possible form. Panel~(b) turns that idea into a small spin-glass model,
which serves as an intuitive template for QAOA-type optimization behavior. Panel~(c) shows an
important programming choice: the same physical building blocks can be optimized with many
independent parameters or with a smaller parameter set tied across layers. Panel~(d) reminds us that
even when the physical angle is conceptually the same, the optimizer may see a folded classical
coordinate.

\paragraph{Parameter tying across circuit layers.}
Sharing the same trainable variable across several parts of a model is commonly called parameter tying.
We use this term throughout. Parameter tying is common in machine learning and is often used for parameter
efficiency and inductive bias. Recurrent neural networks apply the same transition weights at many time
steps, the Universal Transformer applies a common block across depth, and ALBERT ties Transformer
parameters between layers to reduce model size~\cite{bengio1994,pascanu2013,dehghani2019,lan2020}.
The number of trainable variables is reduced, while each tied variable influences several stages of a
nonlinear computation. In recurrent networks this repeated influence is a classical source
of difficult gradient propagation, including vanishing and exploding gradients~\cite{bengio1994,pascanu2013}.

In QAOA, the effect is direct. With independent layers, each pair of angles can be adjusted
separately. With parameter tying, changing one angle simultaneously changes several physical applications of
the same cost or mixer block. The classical search dimension stays small, but the optimizer loses
layer-specific adjustment directions, and the effects of a few variables are coupled across a deeper circuit.
That restriction can reshape the low-dimensional objective and can increase both repeated local structure
and differences among local minima, as the controlled examples below show. The effect depends on the
model. Multi-angle QAOA, for example, relaxes parameter tying and can improve approximation quality at the cost of a larger classical search space~\cite{herrman2022}. Parameter transfer is a
different practice: good values initialize related problems or sizes, while parameter tying enforces equal
parameters inside one circuit~\cite{mele2022,lyngfelt2025}.

Parameter tying and nonlinear reparameterization are different mechanisms. Tying makes one variable
control several circuit locations. Reparameterization changes the map from an optimizer variable to the
executed angle. Both can reshape the objective function. The next
figure isolates the second effect.

\begin{figure}[!htbp]
    \centering
    \includegraphics[width=0.98\linewidth]{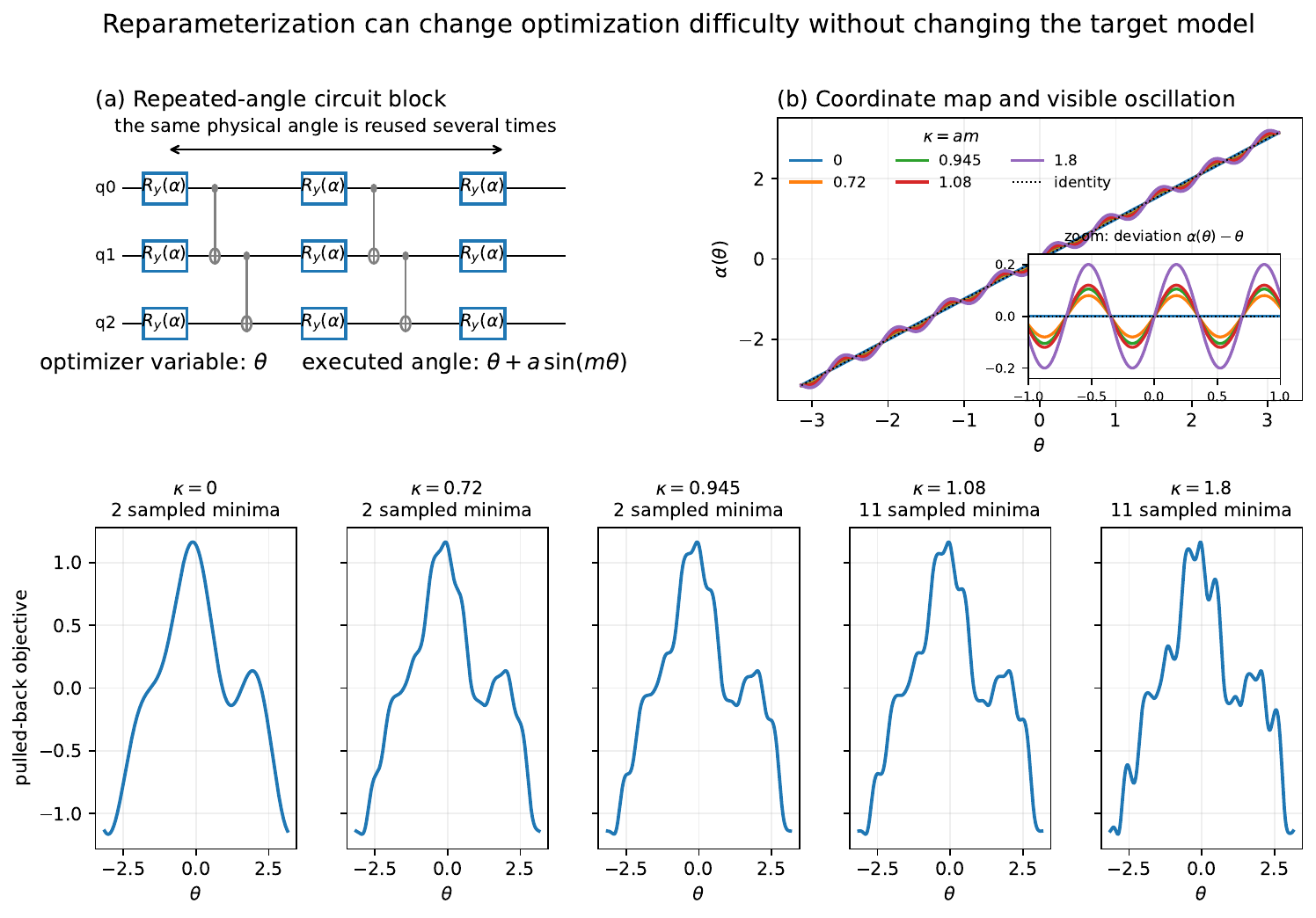}
    \caption{A concrete reparameterization example. (a) A repeated-angle circuit block uses the same
    physical angle $\alpha$ several times. (b) The nonlinear map
    $\alpha(\theta)=\theta+a\sin(m\theta)$ can stay visually close to the identity while introducing a
    high-frequency deviation. (c) After pullback through this map, the same one-dimensional objective
    can develop additional sampled minima.}
    \label{fig:reparam}
\end{figure}

Figure~\ref{fig:reparam} shows the reparameterization directly. A nonlinear map between the
optimizer variable and the executed circuit angle can make a simple periodic objective appear more
multimodal in the optimizer coordinates. We use this example only to illustrate the effect of the
coordinate choice; it is separate from parameter tying.

These design choices show how the objective function can acquire difficult local structure.
The next step is to visualize that structure while keeping the limitations of each projection clear.

\section{Visualization and diagnostics of the objective function}
\label{sec:workflow}

High-dimensional variational objectives require several complementary views. Literal restrictions show
chosen coordinates, optimizer-informed restrictions show directions that matter to search, and
basin-focused summaries describe the organization of sampled minima. Classical test functions provide
a familiar visual reference. We use these views to judge whether multistart local search is likely to be
adequate or whether broader exploration is warranted.

\subsection{Exact low-dimensional restrictions}
The most literal visualization is an exact coordinate slice: choose one or two parameters, hold the rest
fixed, and evaluate the objective on a grid. This is easy to explain and has the virtue of being an
actual restriction of the original function. Its limitation is equally clear: if the chosen coordinates
miss the relevant directions, the picture may look misleadingly simple.

\subsection{Optimizer-informed restrictions}
A more informative restriction chooses its directions from optimization data, for example from the line
between a good solution and a poorer local-search endpoint. Such a view is still exact on the chosen
plane, but the plane itself is selected to expose a practically important part of the geometry.

\begin{figure}[!htbp]
    \centering
    \begin{minipage}[t]{0.49\linewidth}
        \centering
        \includegraphics[width=\linewidth]{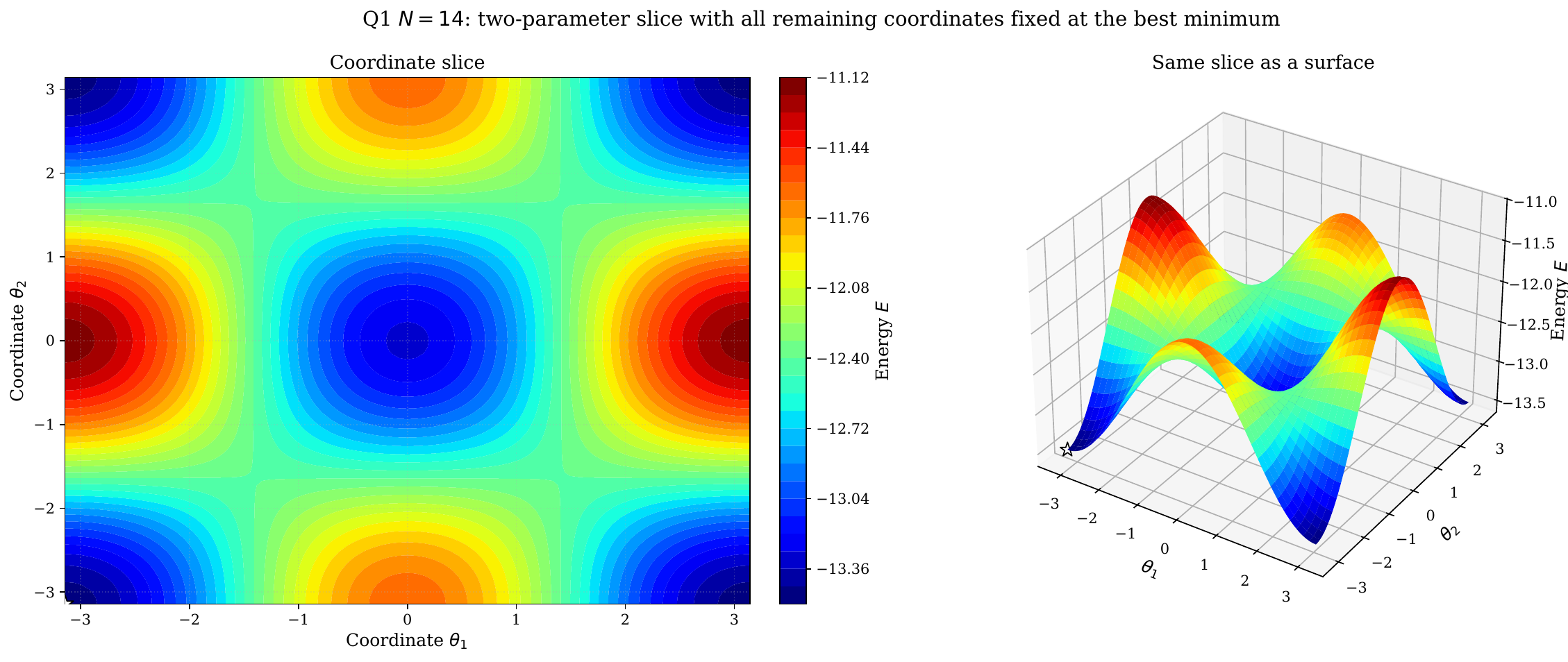}\\[-1mm]
        \small (a) Coordinate slice.
    \end{minipage}\hfill
    \begin{minipage}[t]{0.49\linewidth}
        \centering
        \includegraphics[width=\linewidth]{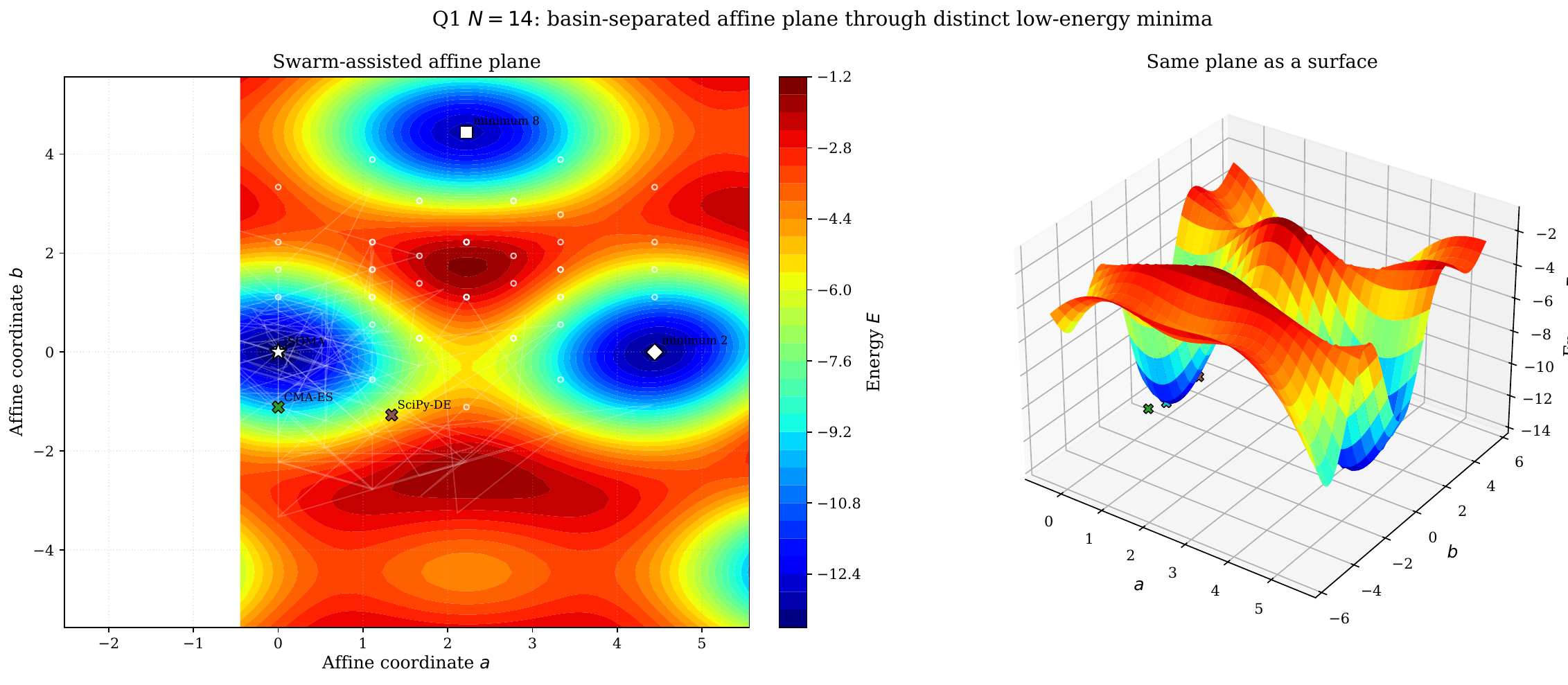}\\[-1mm]
        \small (b) Optimizer-informed plane.
    \end{minipage}
    \caption{Two exact two-dimensional restrictions of the same $N=14$ frustrated-Ising VQE
    objective. The coordinate slice is easy to compute, while the optimizer-informed plane exposes a
    more decision-relevant direction.}
    \label{fig:slices}
\end{figure}

Figure~\ref{fig:slices} illustrates the difference. The first panel answers ``what happens if I literally
plot two parameters?''  The second answers ``what happens along a direction that matters for
optimization?''  They answer different questions. In the optimizer-informed examples used
later, a high-quality anchor is paired with a distinctly poorer local endpoint; their difference defines
one slice direction and an orthogonal companion defines the displayed plane. This is a deliberately selected view of an optimization-relevant direction; it does not summarize the
full high-dimensional objective.

\subsection{Basin-focused summaries}
\begin{figure}[!htbp]
    \centering
    \begin{minipage}[t]{0.48\linewidth}
        \centering
        \includegraphics[width=\linewidth]{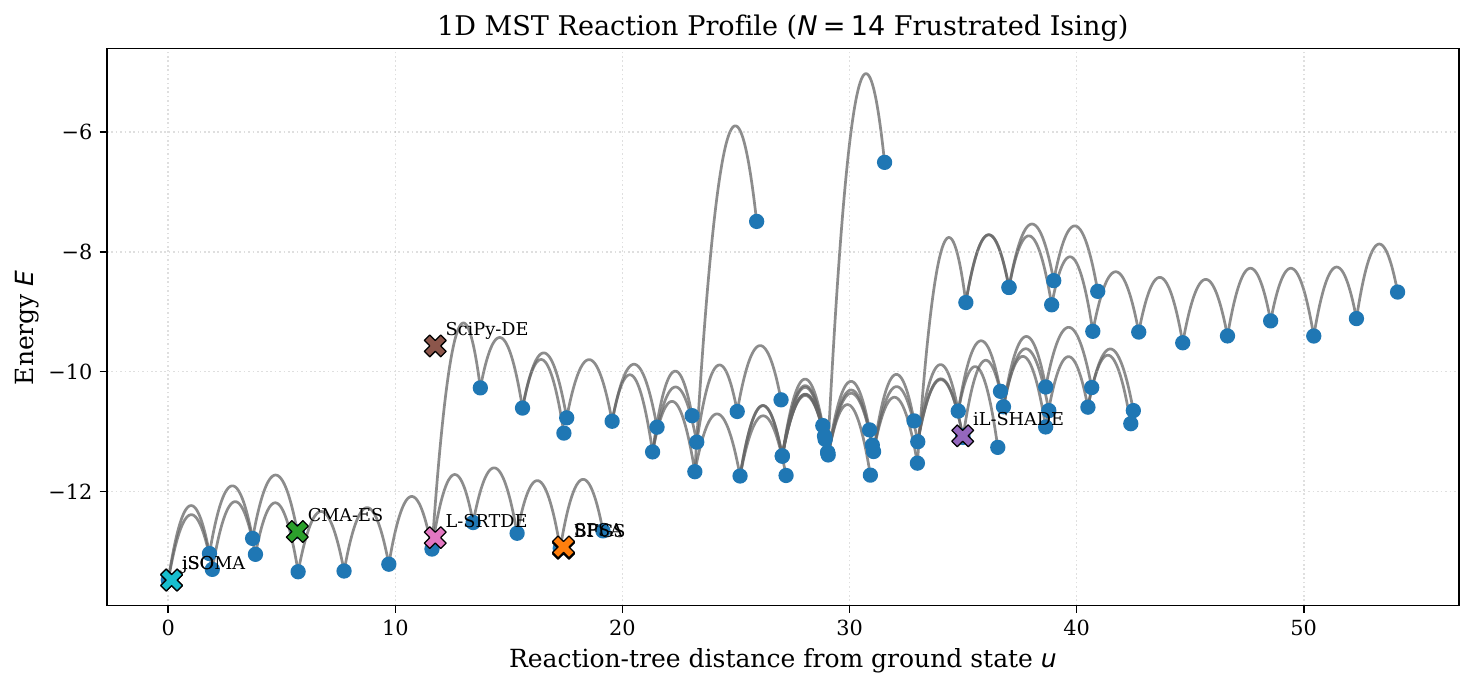}\\[-1mm]
        \small (a) Basin reaction profile.
    \end{minipage}\hfill
    \begin{minipage}[t]{0.50\linewidth}
        \centering
        \includegraphics[width=\linewidth]{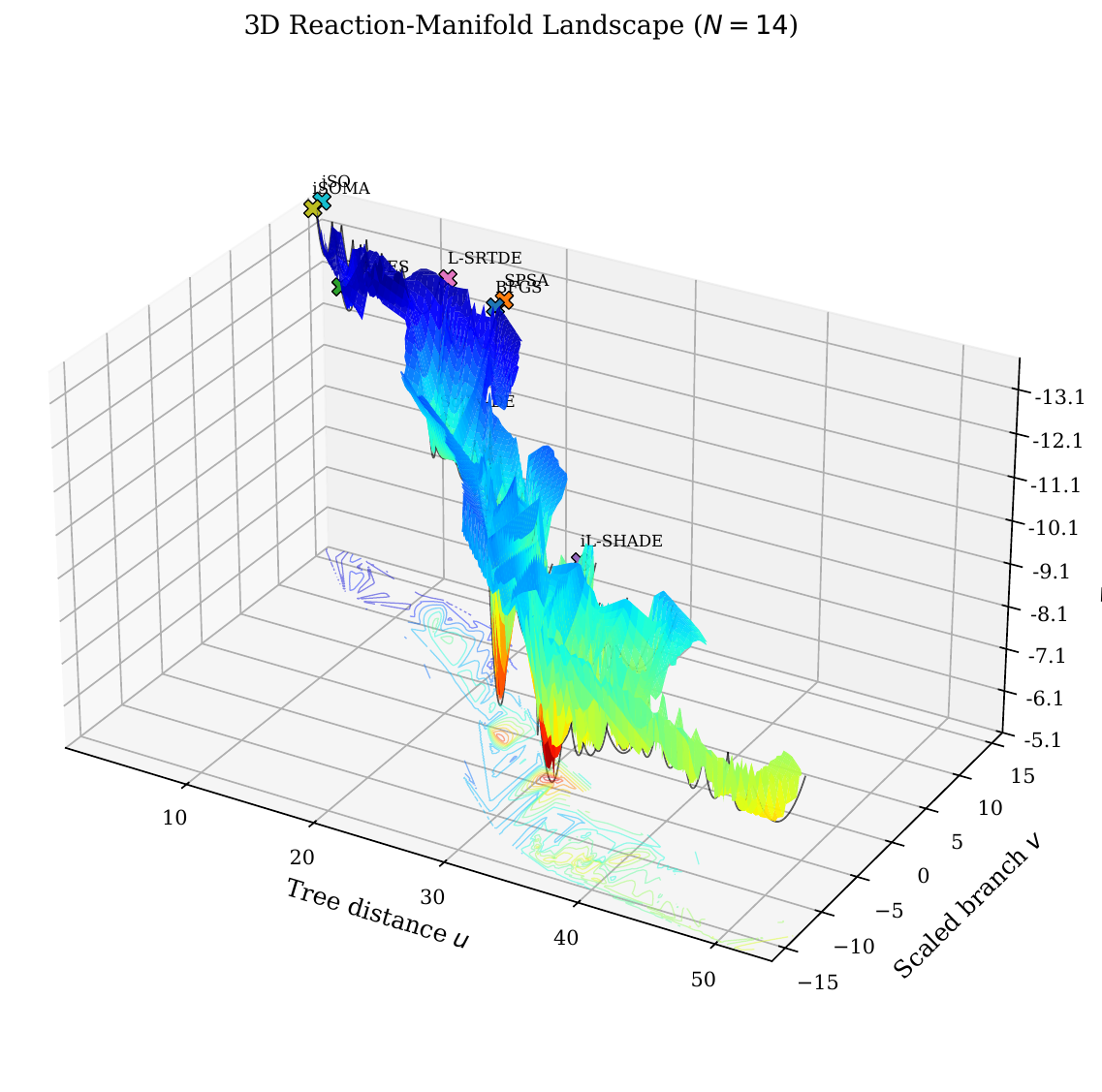}\\[-1mm]
        \small (b) Basin-focused topographic rendering.
    \end{minipage}

    \vspace{1mm}
    \includegraphics[width=0.64\linewidth]{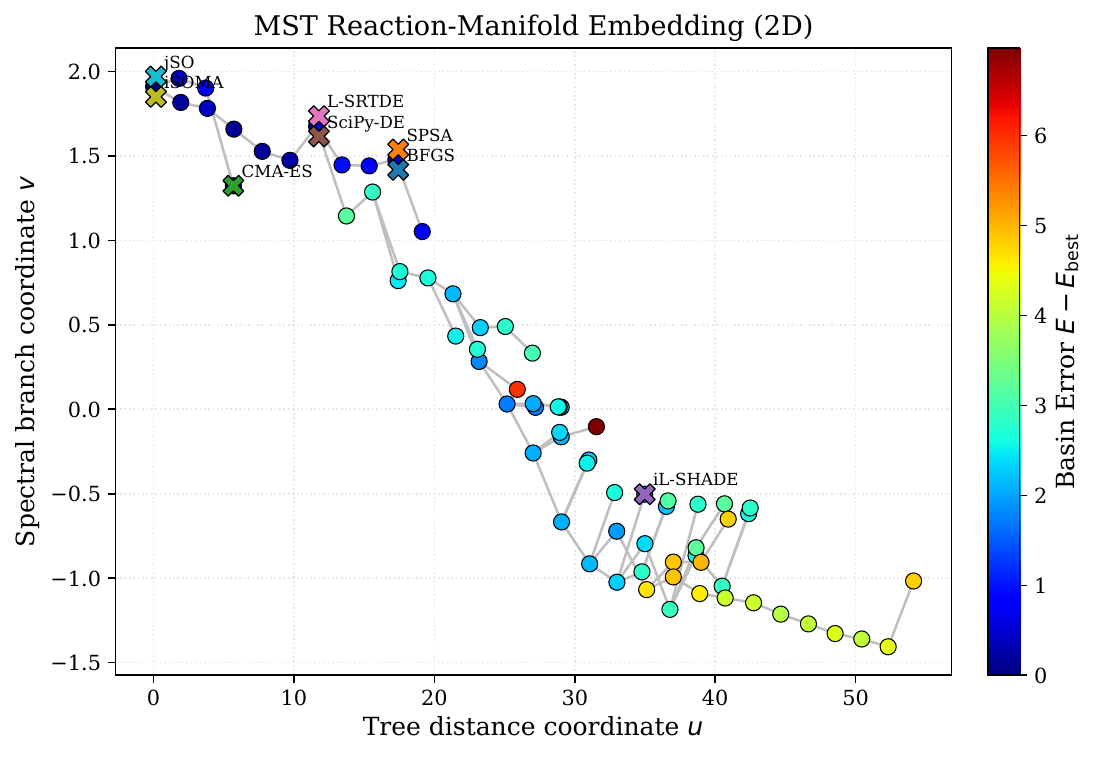}\\[-1mm]
    \small (c) Two-dimensional basin-graph embedding.
    \caption{Three basin-focused views of the same 14-spin frustrated-Ising example. Energies and
    sampled minima are tied to the underlying objective, while panels (b)--(c) use display coordinates constructed from basin organization. These views emphasize global organization that a literal coordinate slice may hide.}
    \label{fig:basin-views}
\end{figure}

For objectives with many distinct basins, literal coordinate plots become less informative. A basin-focused summary in Fig. \ref{fig:basin-views} first samples or identifies minima and
transition structure, then builds display coordinates from that reduced basin graph. The horizontal
axes are no longer original circuit parameters, but the resulting picture can be much easier to read.
The $N=14$ example below is discussed in more detail in the companion VQE study~\cite{novakLandscapeGeometry}.

The example separates basin accessibility from local conditioning. It contains 72 exact one-spin-flip
minima and an effective discrete basin count of about 40.5. The ground-basin capture probability is
$7.32\%$ under discrete descent and $4.03\%$ under continuous random starts, while the median active
Hessian condition number is $3.64$. The local minima are moderately conditioned, but a generic start
has a low probability of reaching the best basin.

\subsection{Classical analogy figures}
For a broad audience it is useful to pair the quantum examples with familiar classical contour
pictures. These provide simple visual references for common nonconvex objective functions.

\begin{figure}[!htbp]
    \centering
    \includegraphics[width=0.95\linewidth]{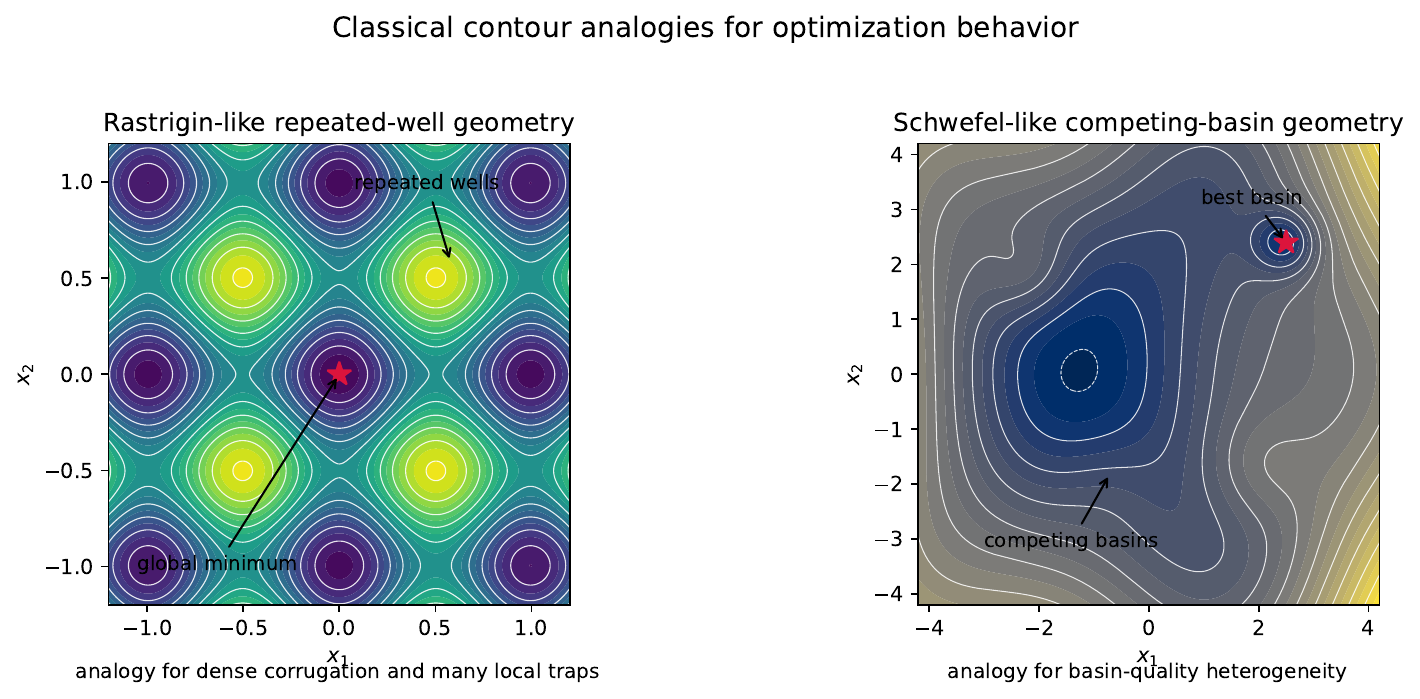}
    \caption{Simple classical analogies. A Rastrigin-like contour shows many repeated local minima.
    A Schwefel-like contour shows minima with substantially different objective values. These contours
    provide a familiar visual reference for the quantum examples.}
    \label{fig:analogies}
\end{figure}

Figure~\ref{fig:analogies} introduces two recurring visual cases. The Rastrigin-like example contains
many local minima, while the Schwefel-like example emphasizes minima with substantially different
objective values.

\section{Objective-function geometry and optimizer behavior}
\label{sec:examples}

The empirical section has three steps. We first define the diagnostics, then compare matched QAOA/VQE
objective functions, and finally test the diagnostics against local and global optimization on MaxCut instances.

\subsection{Empirical protocol and diagnostics}
\label{sec:empirical-conventions}
The empirical results below use exact-objective simulations. Hardware noise is outside the scope of
these examples. For the matched QAOA/VQE examples inherited from the companion studies~\cite{novakGlobalSearch,novakLandscapeGeometry}, objective values are evaluated by exact statevector
simulation. When quantities from different Hamiltonians are compared, energy is converted to the
normalized physical error
\begin{equation}
\epsilon_{\mathrm{phys}}(\boldsymbol{\theta})
= \frac{E(\boldsymbol{\theta})-E_0}{E_{\max}-E_0},
\label{eq:normalized-error}
\end{equation}
where $E_0$ and $E_{\max}$ are the spectral extrema of the Hamiltonian. A difference of $0.04$ in
Eq.~\eqref{eq:normalized-error} is reported as four spectral-range percentage points (pp). Within a
single-instance slice we may display the raw objective instead, because an affine rescaling does not
change the visible location of wells and ridges.

Basin accessibility is estimated with random-start BFGS local minimization from the common bounded
parameter box. Let $e_1,\ldots,e_m$ be the final normalized errors and let
$e_{\min}=\min_i e_i$. Three directly interpretable summaries are
\begin{align}
p_{\mathrm{near}} &= \frac{1}{m}\sum_{i=1}^{m}\mathbf{1}[e_i-e_{\min}\le 0.01], \\
g_{\mathrm{med}} &= \operatorname{median}_i(e_i-e_{\min}), \\
p_{\mathrm{poor}} &= \frac{1}{m}\sum_{i=1}^{m}\mathbf{1}[e_i-e_{\min}>0.05].
\label{eq:basin-diagnostics}
\end{align}
Here $p_{\mathrm{near}}$ measures how often a near-best endpoint is reached,
$g_{\mathrm{med}}$ measures typical endpoint-quality loss, and $p_{\mathrm{poor}}$ measures the frequency
of clearly inferior local outcomes. The companion confirmation and transfer panels use 128 and 64
random local starts, respectively~\cite{novakGlobalSearch}.

Directional multimodality is probed with wrapped one-dimensional parameter-space lines. The simplest
statistic is the number of sampled local minima crossed by a line; Fourier entropy \cite{fefferman1978theory} can additionally
indicate whether variation is concentrated in a few wavelengths or spread across many frequencies. For
the rough-line figures, several random directions are tested from a low-energy anchor and the roughest
line is retained for visualization. These selected slices are descriptive. Aggregate claims use the
random-start and random-line diagnostics across all sampled directions.

Table~\ref{tab:diagnostics} translates these measurements into optimizer-facing language. The first
four are the most directly useful for the local-versus-global question; the local-curvature diagnostic is
included to emphasize that difficult basin organization can coexist with mild conditioning near a
minimum.

\begin{table}[!htbp]
\centering
\small
\caption{Optimization diagnostics and their operational interpretation. Thresholds are expressed in
normalized physical-error units from Eq.~\eqref{eq:normalized-error}.}
\label{tab:diagnostics}
\begin{tabular}{@{}p{0.20\linewidth}p{0.33\linewidth}p{0.37\linewidth}@{}}
\toprule
Diagnostic & Operational definition & What it tells the optimizer \\
\midrule
Minima per line & Mean number of sampled local minima along random wrapped 1D lines & Directional multimodality; many local minima motivate restarts or broader exploration \\
Near-best rate $p_{\mathrm{near}}$ & Fraction of local starts ending within $0.01$ of the best sampled endpoint & Empirical accessibility of a good basin; high values favor local search \\
Median endpoint gap $g_{\mathrm{med}}$ & Median loss of a local endpoint relative to the best sampled endpoint & Variation in local-minimum quality; large gaps make a poor local endpoint costly \\
Poor-endpoint fraction $p_{\mathrm{poor}}$ & Fraction of local starts ending more than $0.05$ above the best sampled endpoint & How often local search is materially misdirected \\
Fourier entropy & Spectral entropy of sampled 1D objective profiles & Whether roughness is dominated by a few wavelengths or distributed across scales \\
Positive-curvature ratio & Ratio of retained positive Hessian eigenvalues near a selected minimum & Local anisotropy/conditioning near the selected minimum; complements basin-accessibility metrics \\
\bottomrule
\end{tabular}
\end{table}

\subsection{Matched QAOA and VQE objective functions}

\begin{figure}[!htbp]
    \centering
    \includegraphics[width=0.96\linewidth]{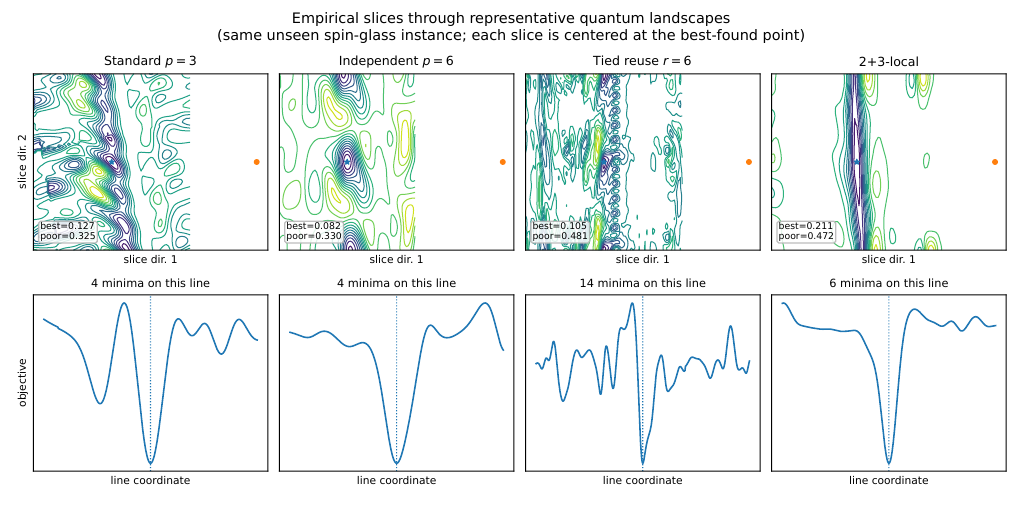}
    \caption{Matched QAOA examples. Independent depth primarily increases repeated local
    structure, while parameter tying and mixed-locality constructions can produce stronger local trapping or
    stronger competition between low-energy regions. The displayed restrictions are informative
    slices, not complete representations of the full objective.}
    \label{fig:qaoa-examples}
\end{figure}

Figure~\ref{fig:qaoa-examples} compares constructions chosen to isolate different effects, following
the controlled comparisons in Ref.~\cite{novakGlobalSearch}. Increasing independent depth adds
oscillatory structure in the displayed restrictions, while local-search endpoint quality changes little.
Parameter tying combines dense repeated structure with poor local endpoints. Mixed locality
produces fewer, more clearly separated low-energy regions. We therefore report the number of sampled
minima and the quality of local-search endpoints separately.

\begin{figure}[!htbp]
    \centering
    \includegraphics[width=0.96\linewidth]{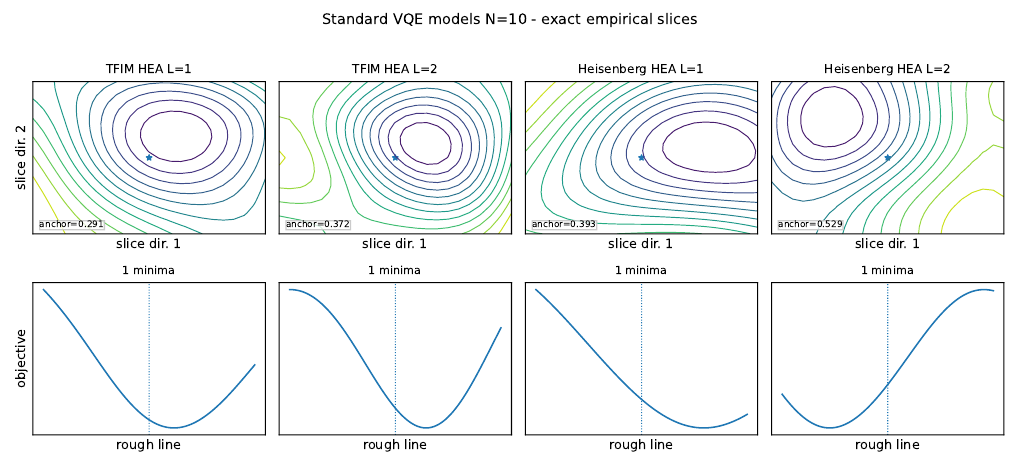}
    \caption{VQE examples on transverse-field Ising model and Heisenberg model. Changes in parameter tying and ansatz structure reshape both the
    number of visible local minima and the spread of low-energy basin quality. Standard physical VQE
    controls can remain comparatively smooth even when related constructed examples are strongly
    multimodal along the sampled directions.}
    \label{fig:vqe-examples}
\end{figure}

Figure~\ref{fig:vqe-examples} shows the same message in a VQE setting. Changes in parameter tying
or ansatz structure can substantially reshape the visible objective. The standard transverse-field Ising model (TFIM) \cite{pfeuty1970one} and Heisenberg model \cite{katsura1962statistical}
controls remain locally accessible in the corresponding study~\cite{novakLandscapeGeometry}.
Related VHA and molecular VQE work also illustrates how ansatz structure, parameter efficiency, and the
targeted state space vary substantially across applications~\cite{illesovaVHA2025,besedaSAOOVQE2024,illesovaQuasiDiabatic2025}.
Optimization difficulty depends on more than compactness, circuit depth, or the fact that the model is quantum.

Additional panels in Appendix~\ref{app:galleries} include analytically controlled Fourier
objectives, spin-glass-inspired rough-line scans, and compact molecular VHA examples.

The matched examples show that the number of local minima and their quality can vary independently. The remaining question
is whether that distinction changes the optimizer that is useful in practice.

\subsection{MaxCut optimizer comparison}
\label{sec:maxcut-showcase}

To test whether these diagnostics correspond to optimizer behavior, we use a small MaxCut experiment. Four unselected random
3-regular graphs with $N=10$ are evaluated with exact statevector objectives. On each graph we compare
standard QAOA at $p=3$, independent depth $p=6$, and a parameter-tied construction in which each of
the three parameterized layers is repeated six times. Thus, the tied construction has 18 physical
alternating layers but retains the same $D=6$ search dimension as standard $p=3$. Each condition uses
64 random-start BFGS quenches and eight wrapped random-line probes. For the optimizer comparison,
FE-matched multistart BFGS (MS-BFGS) and the adaptive differential-evolution method jSO~\cite{brest2017,novakCEC2026} each receive
15,000 function evaluations over eight matched run seeds. The reduced protocol isolates the parameter-tying
construction identified in the companion global-search study~\cite{novakGlobalSearch}. Table~\ref{tab:maxcut-showcase} reports medians across
the four graphs.

\begin{table}[!htbp]
\centering
\small
\caption{Compact MaxCut showcase. Endpoint gaps and poor-endpoint rates use the definitions in
Eq.~\eqref{eq:basin-diagnostics}; ``min./line'' is the median sampled line-minimum count. The last column
is the median per-instance advantage of jSO over MS-BFGS, so positive values favor jSO.}
\label{tab:maxcut-showcase}
\begin{tabular}{@{}lrrrrr@{}}
\toprule
Construction & $D$ & $g_{\mathrm{med}}$ [pp] & $p_{\mathrm{poor}}$ [\%] & Min./line & jSO adv. [pp] \\
\midrule
Standard $p=3$      & 6  & 2.01  & 4.69  & 12.00 & $-0.03$ \\
Independent $p=6$   & 12 & 1.79  & 5.47  & 14.75 & $-4.63$ \\
Parameter tying $r=6$    & 6  & 16.74 & 82.81 & 21.75 & $+1.88$ \\
\bottomrule
\end{tabular}
\end{table}

The independent-depth case is a useful negative control. Relative to standard $p=3$, it increases the
sampled line-minimum count from $12.00$ to $14.75$ without increasing the typical local-endpoint loss
($2.01$ versus $1.79$ pp); at the same budget jSO is worse than MS-BFGS by $4.63$ pp. Parameter tying
changes both aspects at once: the median endpoint gap rises to $16.74$ pp, $82.81\%$ of local starts end
more than five pp above the best local endpoint, and jSO gains $1.88$ pp over MS-BFGS. In this small
controlled example, independent $p=6$ has more sampled minima but no global-search advantage. The tied
condition combines a large endpoint-quality spread with the observed jSO advantage.

\begin{figure}[!htbp]
    \centering
    \includegraphics[width=0.98\linewidth]{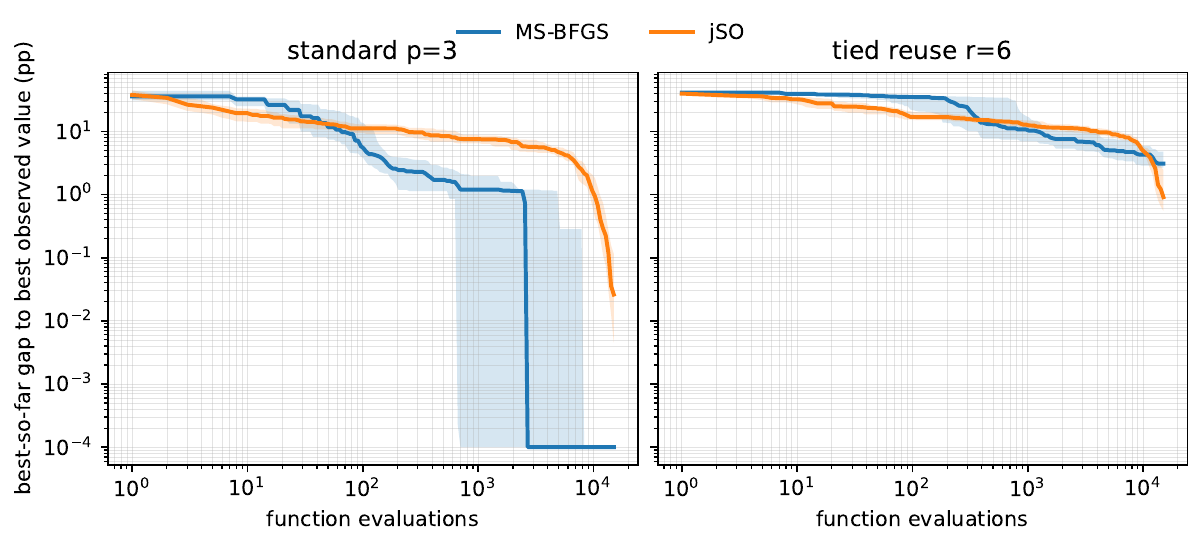}
    \caption{Best-so-far convergence for the two $D=6$ constructions. Curves show the median and
    interquartile range over 32 trajectories per method (four graphs $\times$ eight matched seeds).
    The gap is measured against the best value observed for each graph and construction across the
    local quenches and benchmark runs; it is not a certified global optimum. MS-BFGS rapidly reaches
    the best-observed level for standard $p=3$. With parameter tying, jSO continues to improve late in the
    budget and finishes at lower error.}
    \label{fig:maxcut-convergence}
\end{figure}

Figure~\ref{fig:maxcut-convergence} shows the same result dynamically. For standard $p=3$, MS-BFGS
reaches the best-observed level quickly and jSO catches up near the end of the budget. With parameter tying,
MS-BFGS improves early and then plateaus, while jSO continues to improve and finishes at a lower gap.

\begin{figure}[!htbp]
    \centering
    \includegraphics[width=0.98\linewidth]{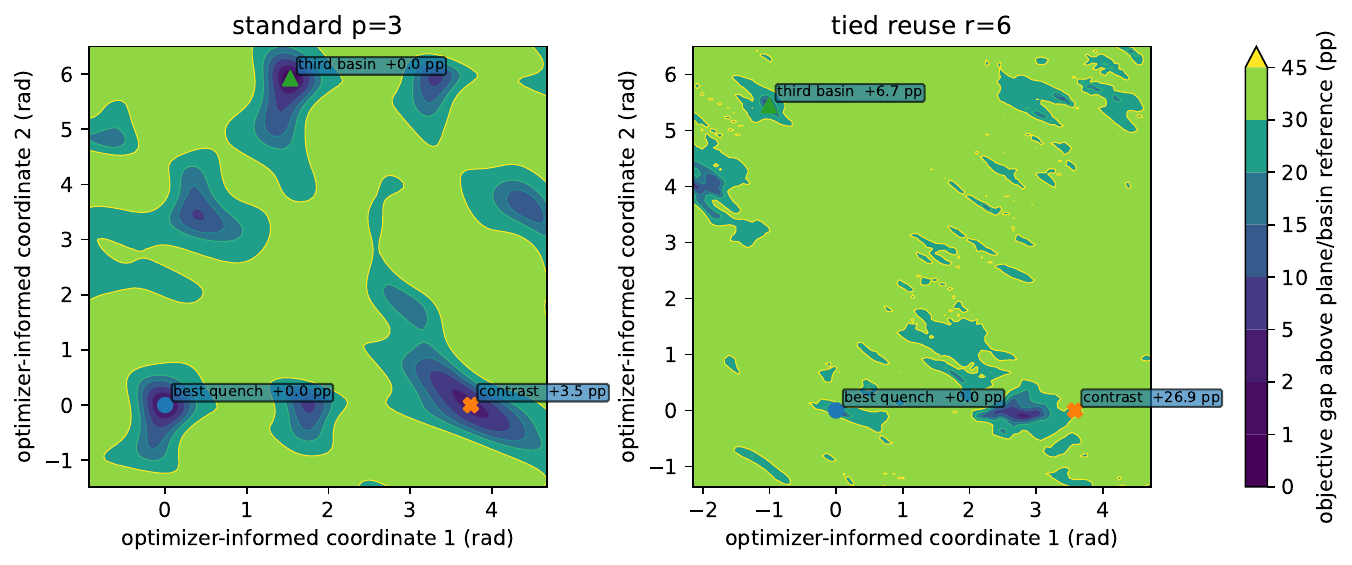}
    \caption{Optimizer-informed two-dimensional restrictions through distinct BFGS endpoints on one
    of the four MaxCut graphs. For standard $p=3$, geometrically distinct endpoints can have
    essentially equivalent quality, and the selected contrast basin is only $3.5$ pp higher. With parameter tying, the same construction exposes substantially unequal basins: the selected third basin is
    $6.7$ pp above the reference and the contrast basin is $26.9$ pp higher. The coordinates are built
    from endpoint-difference directions and are intended to expose basin competition within this plane.
    They do not summarize the full six-dimensional objective function.}
    \label{fig:maxcut-contours}
\end{figure}

The contour view in Fig.~\ref{fig:maxcut-contours} gives the simplest geometric explanation. Standard
$p=3$ can contain several visible local minima with similar objective values. Parameter tying produces local
minima with much larger differences in value. The two panels therefore separate the number of local
minima from the quality of the solutions reached by local search.

\section{Discussion and conclusion}
\label{sec:discussion}

Several practical lessons follow from these results. Multimodality by itself is not a sufficient
description of optimization difficulty: a landscape may contain many local minima without making local
optimization particularly costly if those minima have similar objective values, in which case independent
restarts can provide effective exploration. The situation changes when local optimization terminates in
basins with substantially different objective values, because missing a good basin then becomes costly and
broader exploration can be useful even under a function-evaluation-matched multistart baseline. This
distinction explains the MaxCut results, where increasing independent depth mainly increases sampled local
structure, whereas parameter tying also produces a much larger spread in local-search endpoint quality.

Although the optimizer sees $E(\boldsymbol{\theta})$ simply as a real-valued classical function, its
geometry is induced jointly by the Hamiltonian, the reachable quantum states, and their parameterization.
Different geometric descriptions should therefore not be conflated: the pullback/Fubini--Study metric
measures sensitivity of the represented state to parameter motion, the Hessian describes local curvature
of the scalar objective, and random-start endpoint statistics measure how reliably local optimization
reaches good or poor basins. In the examples considered here, these endpoint statistics are more directly
connected to the local-versus-global optimizer comparison than local curvature alone
~\cite{novakGlobalSearch}. Conversely, the way a population method proposes new points is an
algorithm-side property rather than a property of the objective itself; proposal geometry for SOMA and
differential evolution is studied separately in Ref.~\cite{novakLinearProposal2026}.

Parameter tying provides a particularly simple example of how the optimizer coordinates can matter as
much as their number. Sharing parameters is also common in machine learning, for example in recurrent
or shared-depth architectures, where it can improve parameter efficiency or impose useful inductive
structure~\cite{dehghani2019,lan2020}. The analogy here is limited to the optimization mechanism:
reusing one variable at several stages of a nonlinear computation couples the effect of that variable
across those stages. In the tied QAOA construction, the classical search dimension remains $D=6$, but
each angle controls several physical circuit layers. The resulting problem is therefore lower-dimensional
than an independently parameterized circuit of the same physical depth, but it is not necessarily easier
to optimize. Parameter count alone is consequently a poor proxy for optimization difficulty.

An important limitation of the present experiments is that all objective values are obtained from exact
statevector simulations. The optimizer therefore sees a deterministic expectation value rather than an
estimate constructed from a finite number of quantum measurements. On actual quantum hardware, even an
otherwise ideal circuit produces sampling noise because
$\langle H\rangle$ must be estimated from finitely many measurement shots. This distinction can matter
substantially for optimizer behavior. Quasi-Newton methods such as BFGS rely on sufficiently consistent
local changes in the objective to construct gradient and curvature information and to perform line
searches. Noisy objective differences can therefore make these steps unreliable. Derivative-free
population methods such as differential evolution do not require local gradient estimates and are often
more tolerant of stochastic objective evaluations, although noise can still degrade their selection and
increase the required evaluation budget. Optimization under noisy VQA objectives is studied separately
in Ref.~\cite{novakNoisy2026}. The relative performance observed here for exact objectives should
therefore not be assumed to remain quantitatively unchanged under finite-shot sampling.

Finite sampling is also distinct from quantum-hardware noise. Gate errors, decoherence, calibration
errors, and readout noise can change the objective itself in addition to making its estimate uncertain;
these effects are outside the scope of the present study. Likewise, the examples considered here did not
exhibit an apparent barren-plateau regime: the optimization difficulty studied in this paper comes from
local-minimum and basin structure rather than from an observed collapse of gradient magnitudes. Barren
plateaus are a separate trainability mechanism~\cite{mcclean2018} and may become important for different
ansatzes, depths, or system sizes.

The visualizations in this paper are intended to answer specific optimization questions rather than to
summarize the objective with a single geometric picture. Coordinate and optimizer-informed slices are
exact restrictions of the objective along selected directions, while basin embeddings use constructed
coordinates to show relationships among sampled minima; likewise, minima counted along one-dimensional
probes measure directional multimodality rather than the total number of minima. These views are therefore
most useful when interpreted together with random-start endpoint statistics. The optimizer comparison
should be read with the same scope in mind: it uses small exactly simulated instances, fixed
function-evaluation budgets, and best-observed rather than certified global optima, so it is a controlled
comparison rather than a universal optimizer ranking. Within these experiments, the main distinction is
that increasing sampled local structure alone does not imply an advantage for global evolutionary search;
what matters more is whether local optimization repeatedly reaches solutions of similar quality or becomes
trapped in substantially worse basins. This reflects the broader point that Hamiltonian choice, circuit
construction, and parameterization determine not only which states are reachable, but also the classical
optimization problem presented to the optimizer.

\appendix
\section{Additional examples and controls}
\label{app:galleries}

The following figures in this appendix provide
three complementary controls: analytically constructed objectives that calibrate the visual vocabulary,
spin-glass-inspired scans that isolate parameter tying, and molecular VHA scans that show
that compact parameterizations can remain smooth. They serve as supporting examples.

\begin{figure}[!tbp]
    \centering
    \includegraphics[width=0.8\linewidth]{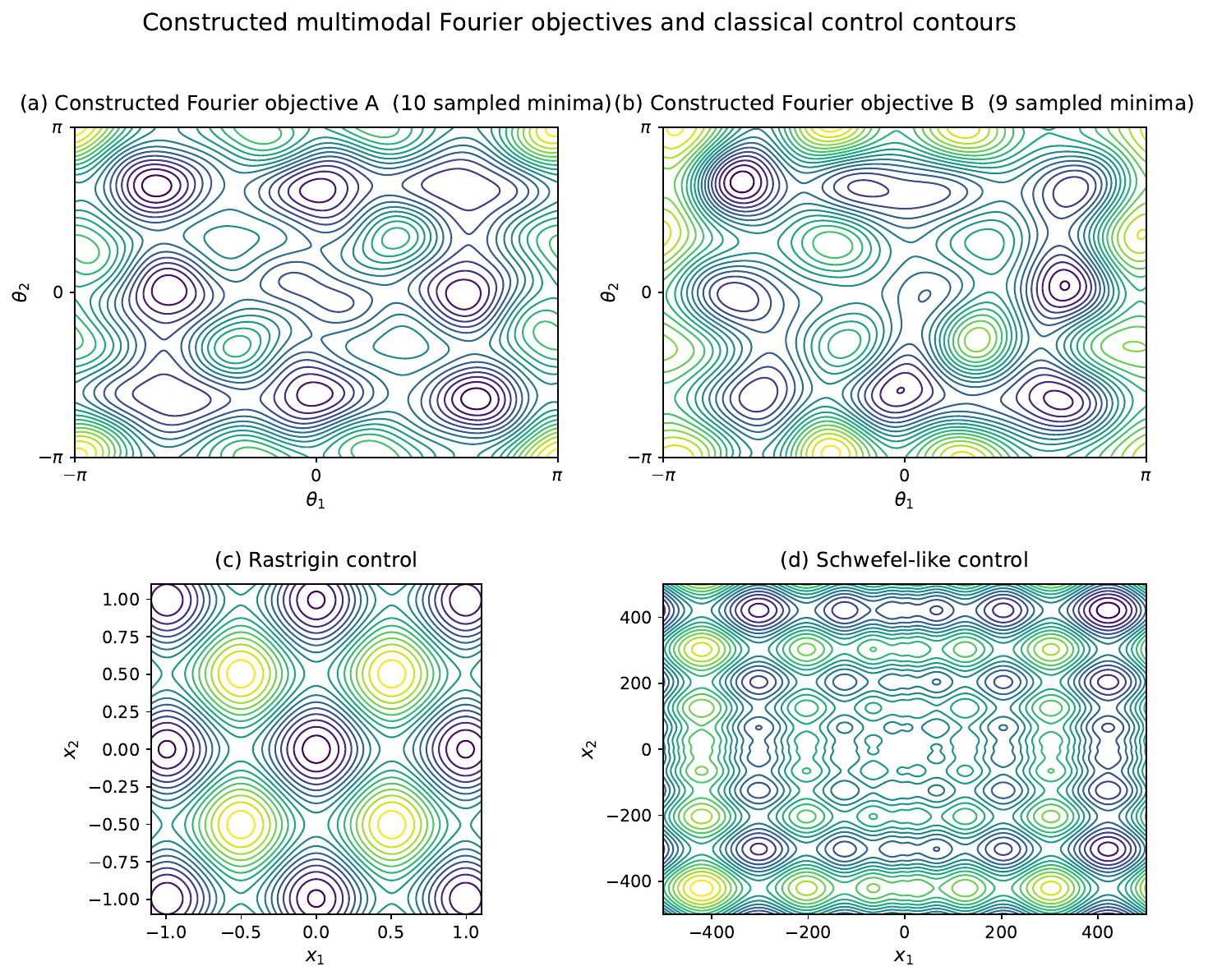}
    \caption{Analytically constructed two-parameter Fourier objectives and classical reference contours.
Panels (a)--(b) provide controlled multimodal examples; panels (c)--(d) show Rastrigin- and
Schwefel-like geometries for visual comparison.}
    \label{fig:fourier-gallery}
\end{figure}

\paragraph{Analytic and classical controls.}
Figure~\ref{fig:fourier-gallery} provides a deliberately simple calibration case. The Fourier
objectives contain several local wells while remaining fully controlled two-parameter functions, so they
show the difference between many local minima and unequal local-minimum quality without any
quantum-model details. The Rastrigin- and Schwefel-like panels connect that distinction to familiar
global-optimization examples.

\begin{figure}[!tbp]
    \centering
    \includegraphics[width=0.9\linewidth]{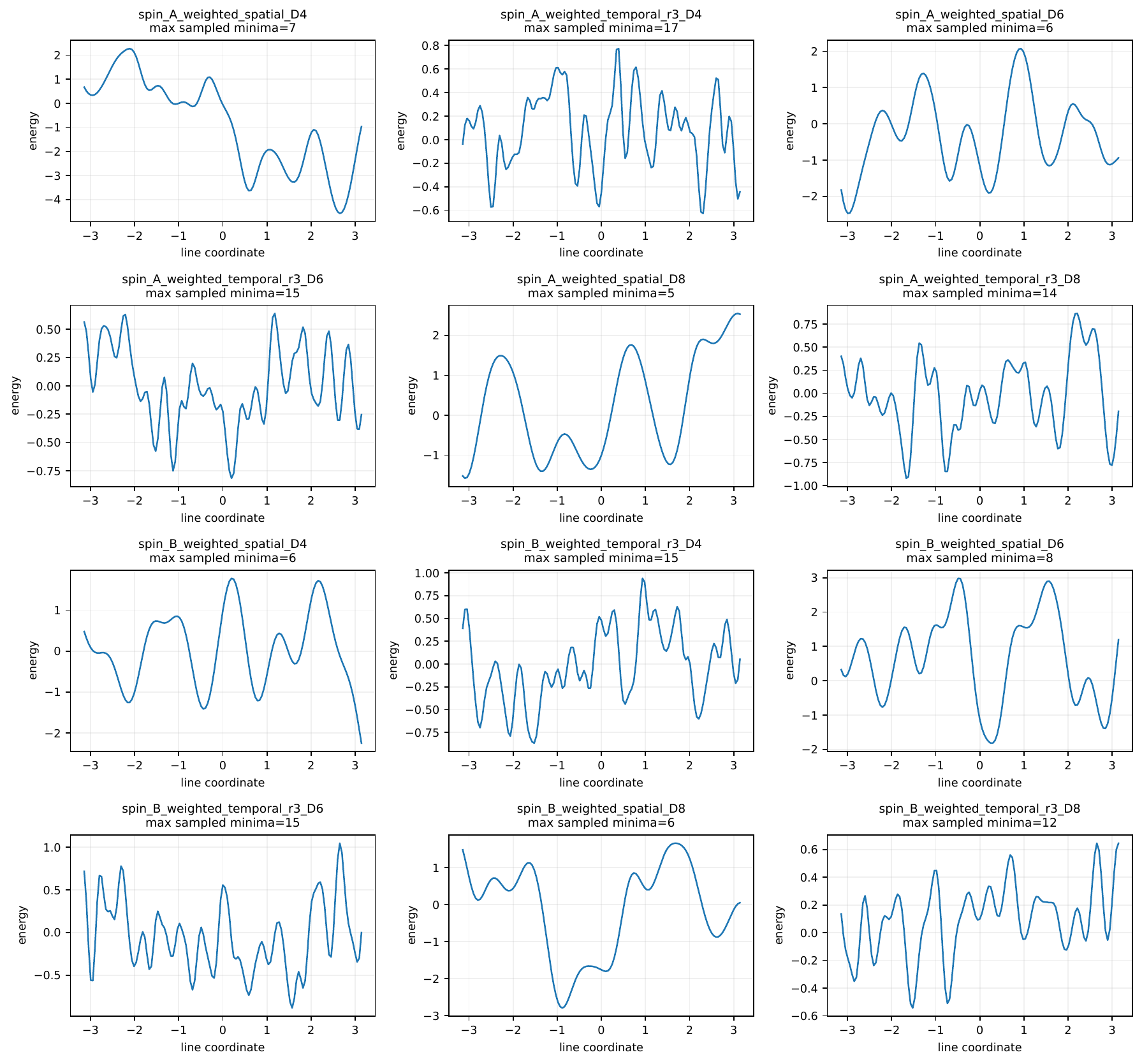}
    \caption{One-dimensional scans through spin-glass-inspired constructions. Spatially weighted and
    parameter-tied variants can have very different sampled minima counts even when built from
    related ingredients.}
    \label{fig:spin-lines}
\end{figure}

\paragraph{Spin-glass parameter-tying controls.}
Figure~\ref{fig:spin-lines} isolates the effect that is most relevant to the QAOA discussion. Across
the displayed dimensions, the parameter-tied variants show substantially more sampled minima along
the selected rough directions than the related spatially weighted variants. Because each panel retains
the roughest of a small set of directions, the figures should be read as an illustrative example; the
aggregate random-line and local-endpoint statistics in the main text carry the quantitative claim.

\begin{figure}[!tbp]
    \centering
    \includegraphics[width=0.9\linewidth]{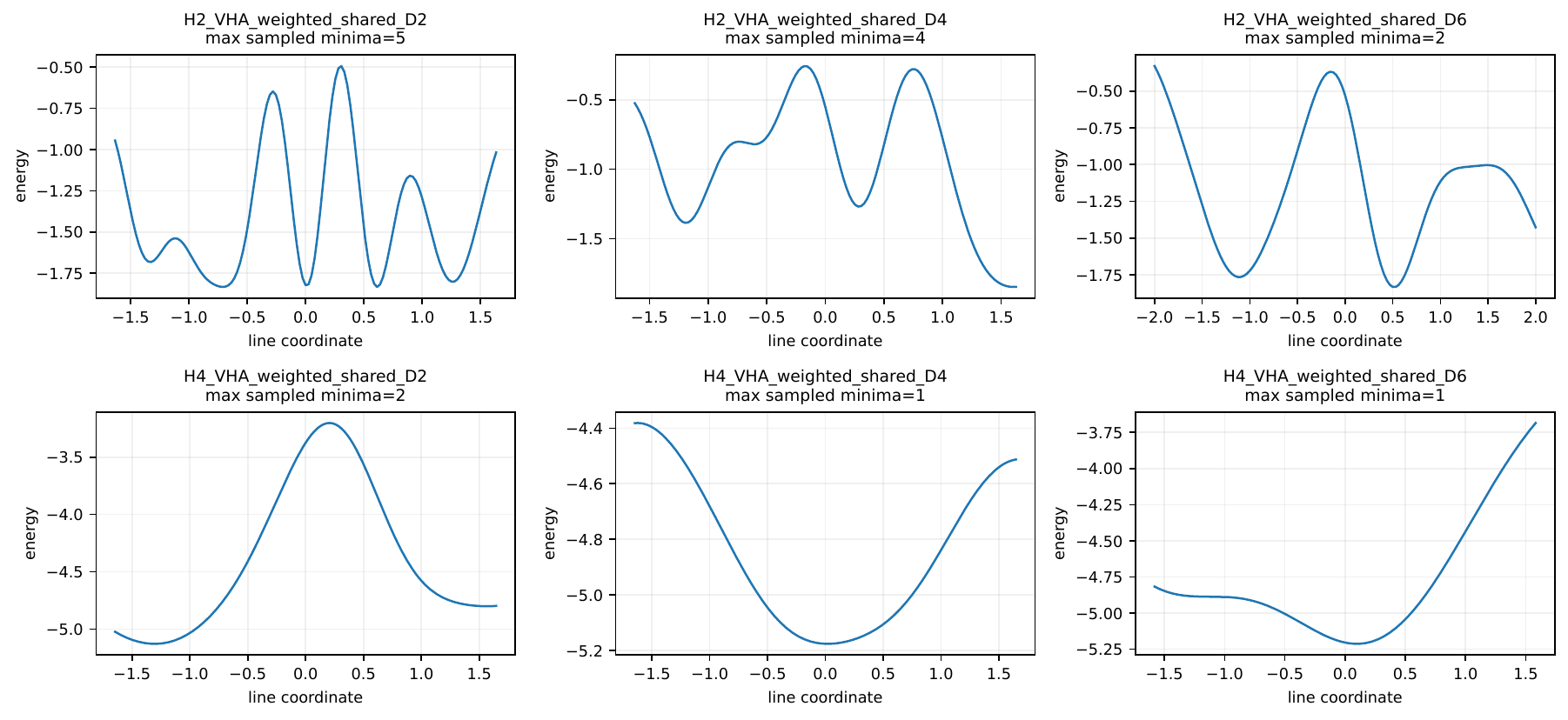}
    \caption{One-dimensional scans through compact molecular VHA constructions.}
    \label{fig:molecule-lines}
\end{figure}

\paragraph{Molecular VHA controls.}
The molecular examples in Fig.~\ref{fig:molecule-lines} provide an important counterexample to any
simple rule that parameter tying or low dimension must produce a difficult objective function. Several of the
H$_2$ and H$_4$ VHA scans remain smooth or contain only a few sampled minima even as the
parameterization changes. These examples show that optimization difficulty depends jointly on the
physical model, ansatz, and parameterization.

Taken together, the appendix galleries support the same interpretation used in the main text. Repeated
structure can be introduced by construction, and its optimization consequence is model dependent.
Useful diagnostics therefore include both a measure of sampled local-minimum density and measures of
the quality and accessibility of solutions reached by local search.

\FloatBarrier
\bibliographystyle{splncs04}
\bibliography{references}

\end{document}